\documentclass[iop,apj,tighten,numberedappendix,twocolappendix,revtex4]{emulateapj}
\usepackage{graphicx}
\usepackage{ulem}
\usepackage{rotating}
\usepackage{lscape}
\usepackage{natbib}
\usepackage[caption=false]{subfig}
\usepackage{float}
\usepackage[pdfa]{hyperref}
\usepackage[capitalize]{cleveref}
\usepackage{filecontents}
\usepackage{paralist}
\usepackage{color}
\hypersetup{colorlinks,
linkcolor=blue,
citecolor=blue,
filecolor=blue,
urlcolor=blue}
\begin{document}

\title{Time-monitoring Observations of Br$\gamma$ Emission from Young Stars}

\author{J.A. Eisner\altaffilmark{1}, G.H. Rieke, M.J. Rieke, K.M. Flaherty\altaffilmark{2}, 
  Jordan M. Stone, T.J. Arnold, S.R. Cortes, E. Cox}
\affil{Steward Observatory, The University of Arizona, 933 N. Cherry
  Ave, Tucson, AZ 85721}
\email{jeisner@email.arizona.edu}

\and

\author{C. Hawkins, A. Cole, S. Zajac, A.L. Rudolph}
\affil{Department of Physics and Astronomy, California State
  Polytechnic University, 3801 W Temple Ave, Pomona, CA 91768}

\altaffiltext{1}{Visiting Fellow, JILA, University of Colorado and
  NIST, Boulder, CO 80309}

\altaffiltext{2}{Current address: Astronomy Department, Wesleyan
  University, Middletown, CT 06459, USA}

\keywords{accretion, accretion discs---infrared: stars---circumstellar
  matter---techniques: spectroscopic---stars: variables: T Tauri,
  Herbig Ae/Be} 


\begin{abstract}
We present multiple epochs of 
near-IR spectroscopy for a sample of 25 young stars, including 
T Tauri, Herbig Ae/Be, and
FU Ori objects.   Using the FSPEC instrument on the Bok 90-inch telescope, we
obtained $K$-band spectra of the 
Br$\gamma$ transition of hydrogen,
with a resolution of $\approx 3500$.  Epochs were taken over a span of
$>$1 year, sampling time-spacings of roughly one day, one month,
and one year.  The majority of our targets show Br$\gamma$ emission,
and in some cases these are the first published detections.
Time-variability is seen in approximately half of the targets showing
Br$\gamma$ emission.  We compare the observed variability with
expectations for rotationally-modulated accretion onto the central
stars and time-variable continuum emission or extinction 
from matter in the inner disk.  Our
observations are not entirely consistent with models of
rotationally-modulated magnetospheric accretion.  Further monitoring,
over a larger number of epochs, will facilitate more quantitative
constraints on variability timescales and amplitudes, and a more
conclusive comparison with theoretical models.
\end{abstract}

\section{INTRODUCTION}
\label{sec:intro}
Young stellar objects have long been known as
variable from UV to IR wavelengths 
\citep[e.g.,][]{JOY42,HERBST+94,SKRUTSKIE+96,CHS01,
EIROA+02,FORBICH+07,FLAHERTY+12}. 
In addition to continuum variability, changes in spectral features
with time have been observed in some sources, with the first
time-monitoring spectroscopy conducted soon after the discovery of T
Tauri stars \citep{JOY45}.   
Subsequent spectroscopic monitoring of T Tauri
stars, predominantly at optical wavelengths, has continued to the
present day, providing constraints on gas kinematics and variability
around young stars.

Variability in the line profile shapes of Balmer transitions, 
for example the appearance or disappearance of inverse P Cygni
profiles, has been linked to changes in winds from, or infall onto, T
Tauri stars \citep[e.g.,][]{BERTOUT+77}. The characteristic
timescale of this variability is often observed to be hours
to days, similar to expectations for magnetically mediated accretion
\citep[e.g.,][]{BBB88,GH93}.  In particular, optical spectroscopic
variability on timescales comparable to stellar rotation periods
has been interpreted as evidence for rotationally-modulated accretion
along stellar magnetic field lines
\citep[e.g.,][]{GIAMPAPA+93,JB95,BOUVIER+07}.
Later studies extended similar monitoring, and results, to the
higher-mass analogs of T Tauri stars, the Herbig Ae/Be stars
\citep[e.g.,][]{MENDIGUTIA+11,COSTIGAN+14}.

Not all young stars exhibit spectroscopic variability
consistent with changes in the magnetospheric accretion process.
In a handful of objects, notably those characterized as UX Ori
variables \citep[e.g.,][]{GRININ+94}, 
spectroscopic variability has been attributed to
time-variable extinction from dusty clouds orbiting in the
inner circumstellar disk
\citep[e.g.,][]{RODGERS+02,SCHISANO+09,RWB11}.
Variability in accretion rate may also lead to changes in inner disk
structure that cause time-variable extinction
\citep[e.g.,][]{STONE+14}.

Observed variability of optical emission lines---notably H$\alpha$ and
other Balmer series lines---is often interpreted in terms of accretion
variability, which is reasonable given the empirical 
correlation of line luminosities with accretion rates
\citep[e.g.,][]{GULLBRING+98,RIGLIACO+12}.
However, optical line profiles for many young stars appear to trace
 a combination of infalling and outflowing matter
\citep[e.g.,][]{RPL96,KHS06}.  This need not destroy the statistical
correlation of optical
line flux and accretion rate, since the outflow rate generally traces
the accretion rate \citep[with a multiplier of about 0.1; e.g.,][]{HEG95}.
However for any given object, it may be hard to distinguish whether
observed variability is due to a variable accretion flow, a
variable outflow, or a combination of both.

One way to distinguish infall and outflow variability is to monitor
Br$\gamma$ emission.  Br$\gamma$ emission is also empirically
correlated with accretion rate \citep{MHC98,MENDIGUTIA+11}.
While H$\alpha$ line profiles often show blue-shifted absorption
components, suggestive of outflowing matter, this absorption is
typically absent in Br$\gamma$ profiles.  Br$\gamma$ lines also
generally show blue-shifted emission centroids.  Overall, the line
profiles of Br$\gamma$ are more consistent with pure infall than
H$\alpha$ and other optical lines \citep[e.g.,][]{NCT96,FE01,MCH01}.  
Br$\gamma$ has the additional advantage that it suffers less
extinction from foreground material; this can be particularly
important for younger, more embedded sources.

A handful of young stars have multiple epochs of Br$\gamma$ emission in
the literature.  For some of these, the epochs were observed with the
same instrument and analyzed in a consistent manner 
\citep{FE01,EISNER+07c,EISNER+10b,SITKO+12,POGODIN+12,
MENDIGUTIA+13,STONE+14}.  Two of these works 
obtained simultaneous
observations of optical and near-IR emission lines, clearly
demonstrating the wind-like line profiles of H$\alpha$ along with the
infall-like line profiles of Br$\gamma$
\citep{POGODIN+12,MENDIGUTIA+13}.  These studies show that the
variability of the optical and near-IR lines are correlated, but
suggest that the variability amplitude may be (marginally)
smaller for Br$\gamma$ emission than for H$\alpha$ emission.  

To better constrain the physical mechanism behind near-IR
spectroscopic variability, we seek monitoring observations over a
range of timescales, for a large, diverse sample of young stars.  
Here we present multi-epoch observations of Br$\gamma$ emission from 
young stars spanning a wide range of both stellar and circumstellar
properties.  This
paper presents a larger sample, with better temporal
coverage, than existing work.  We also present the first observations
of Br$\gamma$ emission from a number of sources, expanding
significantly the existing sample of early-type young stars.

\section{SAMPLE}
\label{sec:sample}
We selected a sample of T Tauri and Herbig Ae/Be stars that were
visible from Tucson during the summer months\footnote{Data for this
program were acquired during the summer to allow undergraduate 
students (co-authors on
the paper) to participate in the observations on Kitt Peak.}.  The sample includes:
5 T Tauri stars, pre-main-sequence analogs of solar-type stars; 
17 Herbig Ae/Be stars, which are young, intermediate mass stars
surrounded by circumstellar material;  2 FU Orionis stars, thought to be
young stars surrounded by particularly active accretion disks; 
and one  heavily-veiled object whose spectral type is uncertain.  We
provide some basic information about our sample below.

The T Tauri stars in our sample are AS 205 N, V1002 Sco, V2508
Oph, AS 209, and V521 Cyg.  The first four are in the Ophiucus or
Upper Sco regions, all at an assumed distance of 160 pc
\citep[e.g.,][]{CHINI81,WALTER+94}.   V521 Cyg is
located in the ``Gulf of Mexico'' region \citep{HERBIG58,ARMOND+11},
at an estimated distance of $\sim 520$ pc \citep{LAUGALYS+06}.
These sources have spectral types between K6 and K0
\citep{EISNER+05,WALTER+94,TORRES+06,HERBIG58}, 
corresponding to stellar masses of $\sim 0.9$--1.5 M$_{\odot}$ 
\citep[e.g.,][]{EISNER+05,HEG95}.  


The Herbig Ae/Be stars in our sample include 6 Herbig Ae stars: HD 141569,
MWC 863, 51 Oph, MWC 275, VV Ser, and V2020 Cyg; and 11 Herbig
Be stars: EU Ser, MWC 297, V1685 Cyg, V1972 Cyg,  AS 442, V2019 Cyg, LkH$\alpha$
169, V645 Cyg, V380 Cep, V361 Cep, and MWC 1080.  These sources have
spectral types between A2 and O7 \citep[e.g.,][]{HILLENBRAND+92,MORA+01,COHEN77}, 
corresponding to a stellar mass range of $\sim 2$--10 M$_{\odot}$
\citep[e.g.,][]{PS93}.     These targets are located in several
different regions, at distances between $\sim 100$ and 3500 pc 
\citep{HERBIG58,STROM+72,CK81,GOODRICH86,DL+91,HILLENBRAND+92,PERRYMAN+97,
MC99,LS02,STRAIZYS+14}.

We included two FU Orionis objects in our sample: V1057 Cyg and V1515
Cyg. These are two of the prototypical FU Orionis objects
\citep{HERBIG66}, and are thought to represent circumstellar disks in the midst of
enhanced accretion events \citep[e.g.,][]{HK96}.  V1057
Cyg is in the North American nebula at a distance of 600 pc
\citep{LS02}, while V1515 is found in the Cyg R1 region at $\sim 1000$
pc \citep{RACINE68}.  

V1331 Cyg is a heavily veiled object lacking clear stellar
photospheric absorption features \citep[e.g.,][]{EISNER+07c}, and
hence its spectral type is unknown.  However it is known to be an
actively accreting object, as traced by strong Br$\gamma$ emission.
V1331 Cyg is one of the few young stars known to exhibit 
strong CO overtone emission \citep[e.g.,][]{CARR89}, perhaps
indicating a particularly active accretion flow for this source
\citep[e.g.,][]{EHS14}.  V1331 Cyg is found in the NGC 7000 region, at
a distance of $\sim 700$ pc \citep[e.g.,][]{HERBIG58,CK81}.


Most objects in our sample have existing Br$\gamma$ spectra in
the literature.  We compare these existing data with our measurements
below in Section \ref{sec:results}.  We also present the
first observations of Br$\gamma$ emission in 10 sources: 2 T Tauri
stars (AS 209 and V521 Cyg); one Herbig Ae star  (V2020
Cyg); and 7 Herbig Be stars (EU Ser, V1972 Cyg, V2019 Cyg, 
LkH$\alpha$ 169, V645 Cyg, V380 Cep, and V361 Cep).  
These Br$\gamma$ spectra of Herbig Be stars result in a sample of 
early-type young stars approximately
twice the size of previous work \citep[e.g.,][]{GARCIALOPEZ+06}.

\section{OBSERVATIONS AND DATA REDUCTION}
\label{sec:obs}
We used the near-IR spectrograph FSPEC at the Bok 90-inch telescope to
monitor the Br$\gamma$ emission around our sample of young stars
(Section \ref{sec:sample}).  We employed a grating with 600 grooves per mm,
providing a resolving power of
$\lambda/\Delta \lambda = 7400$ (per pixel).
FSPEC provides a $2.4'' \times 96''$ slit, with $1.2''$ pixels.  The
slit is $\sim 2$ pixels wide, and the actual spectral resolution is
closer to 3500.

In our observations we nodded along the slit, observing each object
at 5--6 distinct slit positions.  Total integration times (including
all slit pointings) for each observation are listed in Table \ref{tab:sample}.
Observations of target objects were interleaved with observations of
telluric calibrators.  We selected dwarf calibrators with spectral types
later than G6V; in fact all calibrators except HD 174719---used to
calibrate targets in Serpens---have spectral types later than G8V.
Such late-type calibrator stars do not show photospheric 
Br$\gamma$ absorption, so no spurious signals are
introduced into the target spectra.

We developed an IDL-based data reduction pipeline to produce
calibrated spectra \citep[described in detail in][]{EISNER+13}.  
The pipeline works on an entire night of data as
a block, producing telluric-corrected, wavelength-calibrated
spectra for each object observed during the night.  Because we work
with the entire night of data at once, we can use a weighted sum of
all the data to produce higher signal-to-noise in certain calibrations.  

The data reduction procedure includes: creation of median flat and
dark images; generation of a bad pixel mask; correction of bad pixels
in the data; sky subtraction;  shifting-and-adding nod sets; spectral
extraction;  telluric calibration; and wavelength calibration.
Our observations were not taken under photometric conditions, so we
do not attempt to associate real continuum fluxes with our reduced
spectra.  Rather, we divide each spectrum  by its continuum level so
that all spectra for a given object share a common normalization. 
We thus ignore any
potential variability in the continuum flux of our targets.  Such
variations would not be surprising given the known infrared
variability of many young stars \citep[e.g.,][]{SKRUTSKIE+96,EIROA+02}.

\epsscale{0.95}
\begin{figure*}[tbh]
\plotone{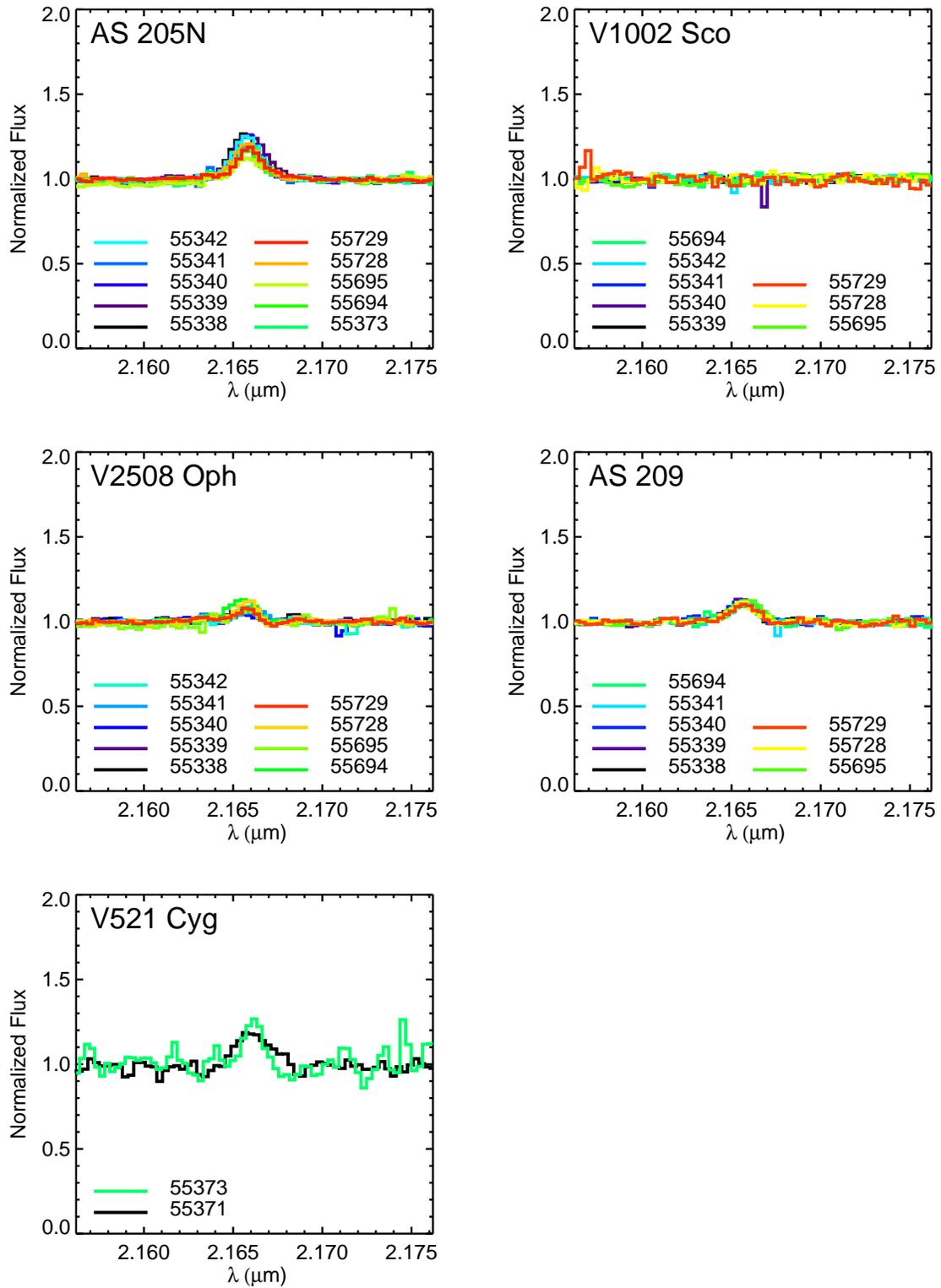}
\caption{Spectra of the T Tauri stars in our sample.  Observed epochs
  for each object are color-coded by MJD.
\label{fig:tts}}
\end{figure*}

\section{RESULTS}
 \label{sec:results}
\begin{figure*}
\plotone{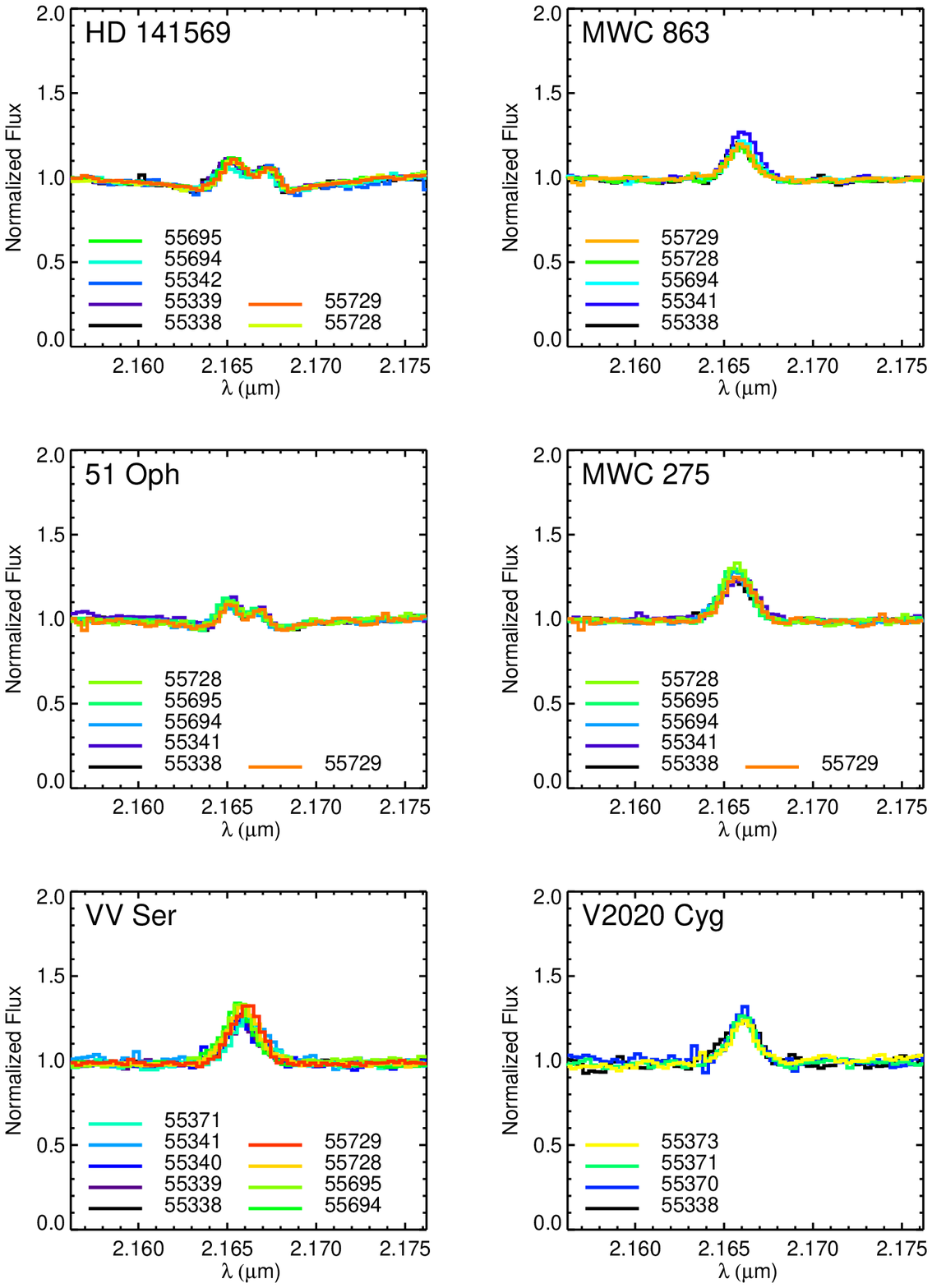}
\caption{Spectra of the Herbig Ae stars in our sample.  Observed epochs
  for each object are color-coded by MJD.
\label{fig:haes}}
\end{figure*}

Spectra of our targets are plotted in Figures
\ref{fig:tts}--\ref{fig:v1331}.  
Nearly all targets show
Br$\gamma$ emission.  A few targets show absorption features, while
two show no discernible Br$\gamma$ features. A number of objects show 
differences in spectra from epoch-to-epoch.

The random uncertainties in each
spectrum can be judged from the channel-to-channel variations in
continuum regions.  While the uncertainties are different from epoch
to epoch, typical error bars are $<1\%$ of the continuum level.  Even
for the faintest objects in our sample (EU Ser and V521 Cyg), the
uncertainties are $<5\%$ of the continuum level.

Because we do not calibrate the continuum flux level of our
observations, we can not calculate Br$\gamma$ line fluxes for our targets.
Rather we determine equivalent widths (EWs), which provide a
measurement of line flux relative to the continuum level.  Although
some fraction of inner disk continuum emission may trace gas, perhaps
free-free emission from H or H$^{-}$
\citep[e.g.,][]{TANNIRKULAM+08,EISNER+09}, 
to first approximation, the continuum emission in the near-IR traces
dust.  Within this approximation, one can view
EW as a measure of Br$\gamma$ gas to dust emission.  

After subtracting the continuum level,
EWs are measured by
simply integrating the flux across the Br$\gamma$ feature, 
from 2.159 to 2.174 $\mu$m.  We do not attempt to correct the spectra
for any photospheric absorption, and so measured EWs may include both
stellar absorption and circumstellar emission (or absorption).
To estimate errors in measured EWs, we simulate 1000 noise
realizations for each spectrum.  Each simulated spectrum consists of
the observed spectrum plus Gaussian  noise, where the $\sigma$
of the noise is determined from line-free regions of the observed spectrum.  
In addition, we include the effects of bad pixels at random locations,
multiplying two pixel values in each synthetic spectrum by 5.  
The uncertainty in the EW measurement is estimated as the 1$\sigma$ confidence
level of the EWs determined for the 1000 synthetic spectra.  
EWs and uncertainties for each target and observed epoch
are listed in Table \ref{tab:sample}.

Two objects in the sample show no evidence of Br$\gamma$ emission (or
absorption) in any observed epoch: V1002 Sco and V361 Cep.  Another
source, V1057 Cyg, shows only hints of Br$\gamma$ features in its
spectra, with most epochs consistent with an EW=0.  Br$\gamma$
emission or absorption features are detected in the remaining 22
targets in our sample.  For 8 objects, these are the
first reported detections of Br$\gamma$ emission: AS
209, EU Ser, V1972 Cyg, V2019 Cyg, V2020 Cyg, V521 Cyg, V645 Cyg,
V380 Cep.  Br$\gamma$ absorption is reported for the first time in
LkH$\alpha$ 169,  although the origin of this spectral feature is
presumably the stellar photosphere rather than circumstellar gas.

About half of the objects with detected Br$\gamma$ emission show
variability, as
traced by measured EWs, across the observed epochs (Table
\ref{tab:sample}).  The sub-sample exhibiting significant Br$\gamma$
variability includes AS 205 N, V2508 Oph, MWC 863, MWC 297, VV Ser,
V1695 Cyg, V1972 Cyg, AS 442, V2019 Cyg, and V1331 Cyg.  Several
objects also show some variability in observed Br$\gamma$ line
profiles from epoch to epoch.  A notable example is V1331 Cyg, where
inverse P Cygni absorption features appear in some epochs (Figure
\ref{fig:v1331}).

In the following subsections, we describe the line profiles and
measured EWs for each target in the sample individually.
We also compare with previous Br$\gamma$ measurements in the
literature, for sources where such data exist.

\subsection{T Tauri Sources}

\subsubsection{AS 205 N}
Our measurements of the EW varied between $-2 \pm 1$ and $-7 \pm 1$
\AA ~over the $>1$ year baseline of the observations.  However, more
modest variations are detected on day-to-day and month-to-month
timescales as well (Table \ref{tab:sample}).  
Previous observations determined EWs of -4.7 \AA
~in 1994 \citep{NCT96}, and -3.7 \AA ~around 2009 \citep{EHS14},
consistent with the range of EW seen in the FSPEC data. 

\subsubsection{V1002 Sco}
We see no Br$\gamma$ absorption or emission features in any observed
epochs.    This contrasts with a previous measurement from 2005, which
showed a broad, asymmetric Br$\gamma$ feature with 
EW = $-14 \pm 6$ \AA $ $ \citep{EISNER+07c}. 

\subsubsection{V2508 Oph}
The Br$\gamma$ emission from V2508 Oph exhibits variability in EW
beyond the uncertainties over the $>1$ year observed time-baseline.
The largest emission feature observed has
an EW of $-4 \pm 1$ \AA, while the weakest emission gives an EW of $0 \pm
1$ \AA  ~(Table \ref{tab:sample}).  This very weak emission feature occurs one
day after, and one month before, stronger features with EWs of $-3 \pm
1$ \AA.  We also see some change in the line profile shapes on the same
timescales, with additional blue emission, or red absorption, in some
epochs (Figure \ref{fig:tts}).
Thus, variability appears to be occurring on short timescales.
However, a previous measurement of Br$\gamma$ emission with an EW of 
-6.5 \AA ~in 2005 \citep{EHS14} suggests that the emission may also
vary on longer timescales.

\subsubsection{AS 209}
We detect Br$\gamma$ emission from AS 209 for the first time.
We see no significant variations across our observed epochs, which
span $>1$ year  (Table \ref{tab:sample}; Figure \ref{fig:tts}).

\subsubsection{V521 Cyg}
V521 Cyg is one of the fainter objects in our sample ($m_K \approx
8.8$), and thus the signal-to-noise of the FSPEC data 
are lower for this object.  However, we clearly detect Br$\gamma$
emission in two separate observations (Figure \ref{fig:tts}).  These
are the first detections of Br$\gamma$ emission in this source.

\subsection{Herbig Ae Sources}

\subsubsection{HD 141569}
The Br$\gamma$ emission from HD 141569 appears to be constant, within
uncertainties, across our observed epochs (Figure \ref{fig:haes}).  
The Br$\gamma$ profiles
include a broad stellar photospheric absorption
component and a double-peaked circumstellar emission component,
consistent with previous observations
\citep{GARCIALOPEZ+06,BRITTAIN+07}.
The EWs listed in Table \ref{tab:sample} include the stellar and
circumstellar components, and are compatible with the earlier
measurements.  For example, 
\citet{GARCIALOPEZ+06} measured an EW of 6 \AA, but if
the stellar absorption is removed, the circumstellar Br$\gamma$ EW is
estimated as -4.5 \AA. 
Since we do not expect the stellar photospheric
absorption to vary, the lack of observed variability in the Br$\gamma$
spectra implies a lack of variability in the cirumstellar component as
well.

\subsubsection{MWC 863}
Measured EWs for MWC 863 range from $-2 \pm 1$ to $-5 \pm 1$ \AA ~over a
$\sim 1$ year timescale (Table \ref{tab:sample}).  However, the
characteristic timescale for variability may be shorter, as suggested by
the marginally significant night-to-night variations.  The line profile
also appears to show additional red emission during the epoch with the
highest measured EW (Figure \ref{fig:haes}).  Previous
measurements of the EW for MWC 863 are between -5.5 and -6.2 \AA
~\citep{GARCIALOPEZ+06,EHS14}, consistent with the range seen here.

\subsubsection{51 Oph}
The line profiles of Br$\gamma$ emission observed for 51 Oph show no
significant changes across the $>1$ year time-baseline (Figure
\ref{fig:haes}), and measured EWs do not vary significantly (Table
\ref{tab:sample}).  As for HD 141569, the EWs include both a broad
stellar absorption component and a double-peaked circumstellar
emission component.  An existing spectrum of 51 Oph in the literature
yields an EW of 1.3 \AA ~\citep{GARCIALOPEZ+06}, consistent with the
measurements presented here.

\subsubsection{MWC 275}
MWC 275 shows no significant variations in Br$\gamma$ EW or line
profiles across the observed epochs (Table \ref{tab:sample}; Figure
\ref{fig:haes}).  Previous determinations of the EW, -6.9 \AA
~\citep{GARCIALOPEZ+06}, -7.1 \AA ~\citep{DB11}, and -6.3 \AA
~\citep{EHS14} are compatible with
our measurements.  However, \citet{MENDIGUTIA+13} measured EWs
between -4.8 \AA ~in October of 2011 and -3.5 \AA ~in February of 2012;
these values are outside the range seen in the present work.  While
the EW estimates and quoted error bars of \citet{MENDIGUTIA+13} imply
variability, the amplitude would fall within our measurement
uncertainties.  \citet{KRAUS+08} also cited evidence for modest line
profile changes, consistent with low-level variability in this object.

\subsubsection{VV Ser}
During the epochs we observed VV Ser, between 22 May 2010 and 16 June
2011, the Br$\gamma$ EW ranged from $-5 \pm 1$ to $-9 \pm 1$ \AA, with
variations apparent on night-to-night timescales (Table
\ref{tab:sample}). Line profiles show slight variations on similar
timescales (Figure \ref{fig:haes}).  While we also observe shifts in
the line centroids, these shifts are $\la 50$ km s$^{-1}$, and may
simply reflect a combination of Earth motion and uncertainties in
the wavelength calibration.
Previous measurements of Br$\gamma$ emission determined EWs of -9.0
\AA ~in 2004 \citep{GARCIALOPEZ+06}, -3.7 \AA ~in 2008 \citep{DB11},
and -5.1 \AA ~in 2011 \citep{EHS14}, generally consistent with the
range we observe here.

\subsubsection{V2020 Cyg}
We detect Br$\gamma$ emission from V2020 Cyg over four epochs, with
three epochs separated by about a day, and the fourth a month before
(Figure \ref{fig:haes}).  To the best of our knowledge this is the
first detection of Br$\gamma$ emission in this object.  No significant
variability is observed (Table \ref{tab:sample}).

\begin{figure*}
\plotone{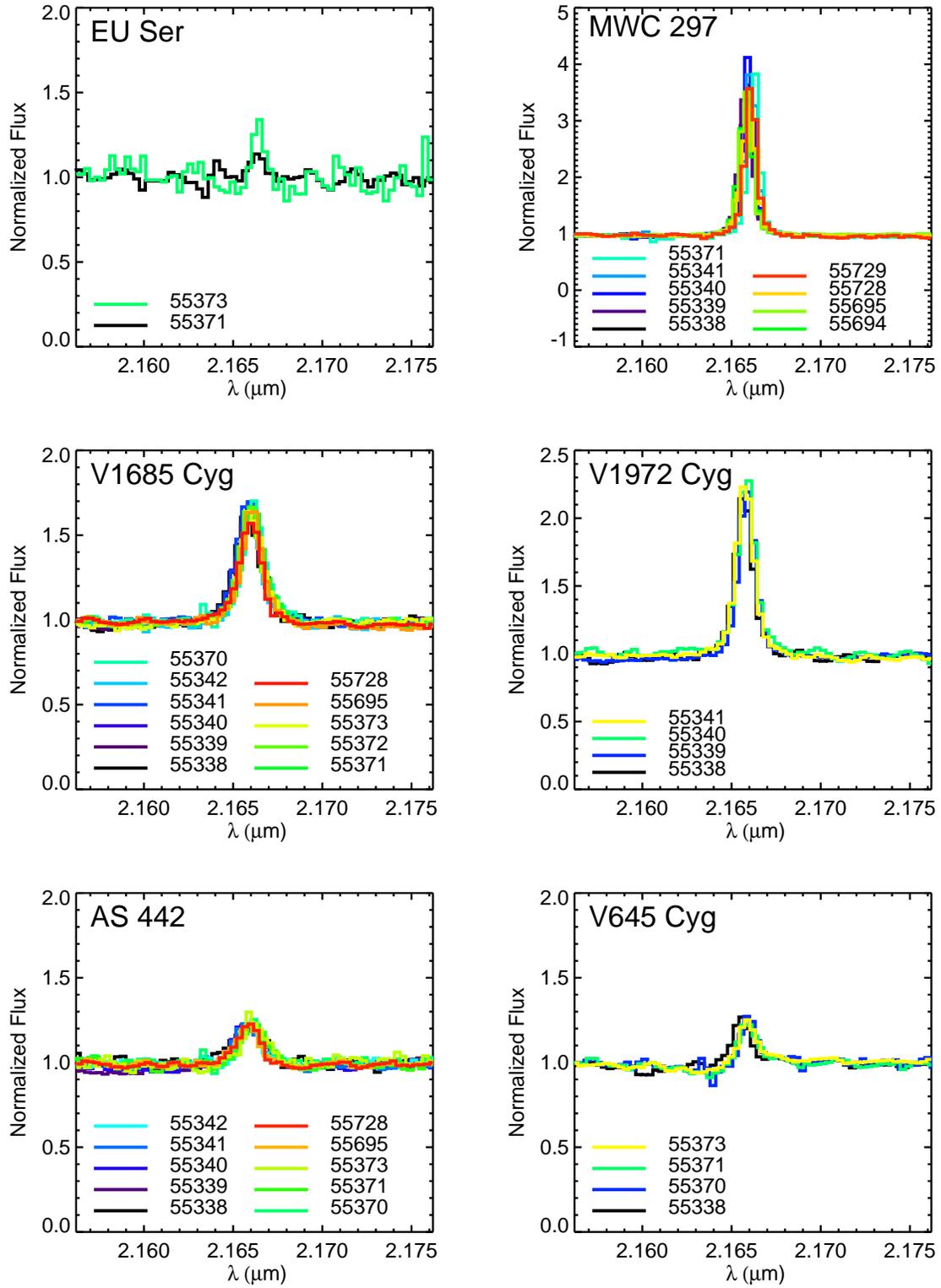}
\caption{Spectra of the Herbig Be stars in our sample.  Observed epochs
  for each object are color-coded by MJD.
\label{fig:hbes}}
\end{figure*}

\begin{figure*}
\ContinuedFloat
\plotone{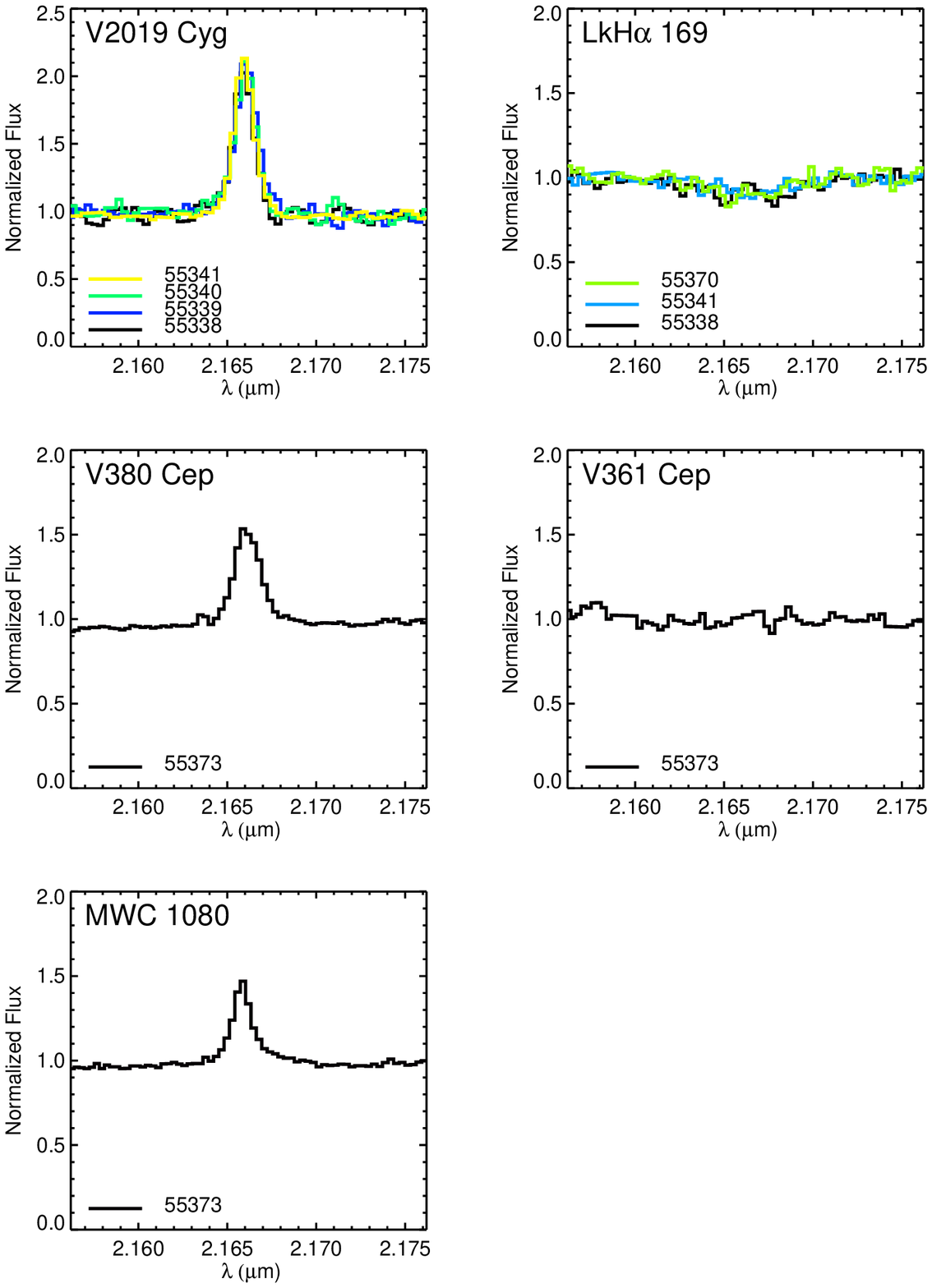}
\caption{continued.
\label{fig:hbes2}}
\end{figure*}

\subsection{Herbig Be Sources}

\subsubsection{EU Ser}
EU Ser is the faintest object in our sample, with $m_K \approx 9.5$.
As illustrated in Figure \ref{fig:hbes}, our measured
spectra are quite noisy.  However, there is clear evidence for the
presence of  Br$\gamma$ emission; this is the first observation of
Br$\gamma$ emission from this source.  We obtained two epochs of FSPEC
data separated by 2 nights, which suggest different line profiles.   
With the low signal-to-noise of the data, however,
this hint of variability is not significant (Table \ref{tab:sample}).

\subsubsection{MWC 297}
MWC 297 is the brightest object in our sample, with $m_K \approx 3$,
and also the target with the largest observed Br$\gamma$ EW (Table
\ref{tab:sample}).  The emission does appear somewhat variable, with EWs
ranging between $-22 \pm 1$ \AA ~and $-28 \pm 1$ \AA ~during the $>1$
year span of our observations.  This large variation is seen on
monthly and yearly timescales, while night-to-night variations are
less significant.  
The line profiles may also differ from
epoch to epoch, although it is difficult to constrain this well, since
the line is fairly narrow for this source (Figure \ref{fig:hbes}).
Our measurements for MWC 297 differ significantly from previous
measurements in the literature: -16.0 \AA ~in 2004 
\citep{GARCIALOPEZ+06}, and -17.2 \AA ~in 2008 \citep{WEIGELT+11}.
Thus, we see evidence for variations on timescales of months and years
in this target.

\subsubsection{V1685 Cyg}
V1685 Cyg shows strong Br$\gamma$ emission with a line profile whose
shape does not vary significantly over 11 epochs observed over the
course of $> 1$ year (Figure \ref{fig:hbes}).  However, the strength
of the emission, as traced by the measured EW, does change.
Variations in EW between $-9 \pm 1$ and $-14 \pm 1$ \AA ~are seen over a
range of timescales (Table \ref{tab:sample}).  A previous measurement of
Br$\gamma$ emission in V1685 Cyg determined an EW of -12.1 \AA
~\citep{EHS14}, within the range of values determined here.

\subsubsection{V1972 Cyg}
We observed V1972 Cyg during four epochs over consecutive days.  We
detected strong Br$\gamma$ emission (Figure \ref{fig:hbes}), the
first published detection for this object.
Significant variations are seen, with EWs
ranging between $-16 \pm 1$ and $-23 \pm 1$ \AA ~(Table \ref{tab:sample}).

\subsubsection{AS 442}
AS 442 shows variability in Br$\gamma$ emission, with measured EWs
between $-3 \pm 1$ and $-7 \pm 1$ \AA ~(Table \ref{tab:sample}).  The
variability occurs on night-to-night timescales.  Line profiles 
also vary slightly from epoch to epoch (Figure \ref{fig:hbes}).  A
previous measurement of
Br$\gamma$ emission in AS 442 determined an EW of -4.5 \AA
~\citep{EHS14}, within the range of values determined here.

\subsubsection{V645 Cyg}
We detect Br$\gamma$ emission from V645 Cyg over four epochs, with
three epochs separated by about a day, and the fourth a month before
(Figure \ref{fig:hbes}).  To the best of our knowledge this is the
first detection of Br$\gamma$ emission in this object.  No significant
variability is observed in the EW (Table \ref{tab:sample}).  However, the spectra
for some epochs suggest absorption on the blue side of the line
(Figure \ref{fig:hbes}), as in P Cygni wind
profiles.  Such a profile is not seen in one spectrum, which was taken
about a year apart from all other epochs.  While this observation
hints at the possibility of a time-variable wind, with the
noise in the data we cannot make a confident statement about variability.

\subsubsection{V2019 Cyg}
We present the first detection of Br$\gamma$ emission toward V2019
Cyg (Figure \ref{fig:hbes}).  
Strong emission was observed during four epochs over consecutive
nights, with EWs ranging between $-17 \pm 1$ and $-21 \pm 1$ \AA ~(Table
\ref{tab:sample}).

\subsubsection{LkH$\alpha$ 169}
We observed LkH$\alpha$ 169 three times, with epochs separated by two
days, and about a month.  No Br$\gamma$ emission is detected, although
we do see broad Br$\gamma$ absorption in all epochs (Figure
\ref{fig:hbes}).  The spectral type of LkH$\alpha$ 169 is B8, and we
expect broad, photospheric Br$\gamma$ absorption.
These spectra are fairly noisy (this object is
relatively faint, with $m_K \approx 8.6$), and the measured EWs are
consistent with no variability.

\subsubsection{V380 Cep}
We present only a single epoch of Br$\gamma$ emission for V380 Cep
(Figure \ref{fig:hbes}; Table \ref{tab:sample}).  This is the first
published Br$\gamma$ emission for this source.
\epsscale{1.0}

\epsscale{1.0}
\begin{figure*}
\plotone{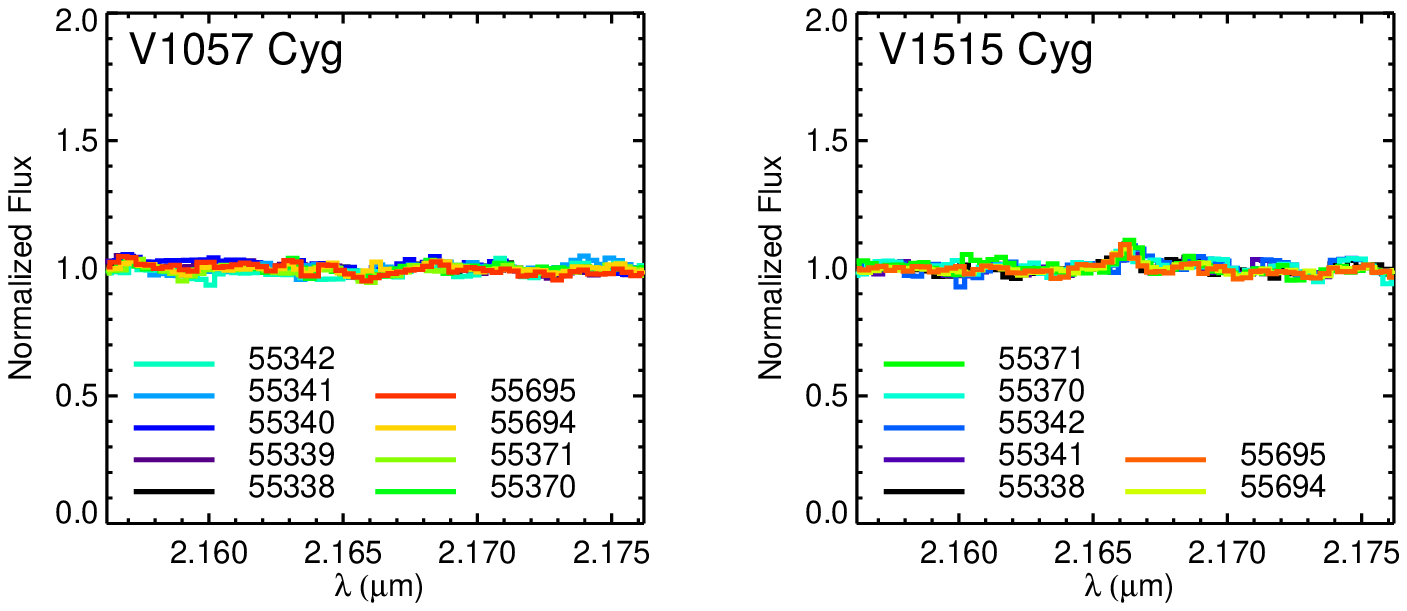}
\caption{Spectra of the FU Ori stars in our sample.  Observed epochs
  for each object are color-coded by MJD.
\label{fig:fuors}}
\end{figure*}

\subsubsection{V361 Cep}
We present only a single observation of  V361 Cep
(Figure \ref{fig:hbes}; Table \ref{tab:sample}).  No Br$\gamma$
absorption or emission is detected for this target. 

\subsubsection{MWC 1080}
We present only a single epoch of Br$\gamma$ emission for MWC 1080,
and measure an EW of $-7 \pm 2$ \AA ~(Figure \ref{fig:hbes}; Table
\ref{tab:sample}).  A previous measurement estimated the Br$\gamma$ EW as
-5.1 \AA ~\citep{EHS14}, consistent with the EW determined here.

\subsection{FU Ori Objects}

\subsubsection{V1057 Cyg}
The spectra for V1057 Cyg, observed over 9 epochs spanning about a
year, show no significant variability (Table \ref{tab:sample}).  Indeed,
measured EWs are consistent with 0 for most epochs, indicating that no
Br$\gamma$ emission or absorption was detected.  However, Figure
\ref{fig:fuors} shows hints of an absorption feature at the expected
wavelength of the Br$\gamma$ transition with some excess red emission, 
perhaps consistent with previous observations \citep{EH11}. 

\subsubsection{V1515 Cyg}
V1515 Cyg appears to show weak Br$\gamma$ emission (Figure
\ref{fig:fuors}), although measured EWs are consistent with 0 for some
observed epochs (Table
\ref{tab:sample}).  No significant variability is observed over the $\sim 1$ year of
our observations.  The line profile appears similar to the one
presented in \citet{EH11}.

\subsection{The heavily-veiled source, V1331 Cyg}
\begin{figure}
\plotone{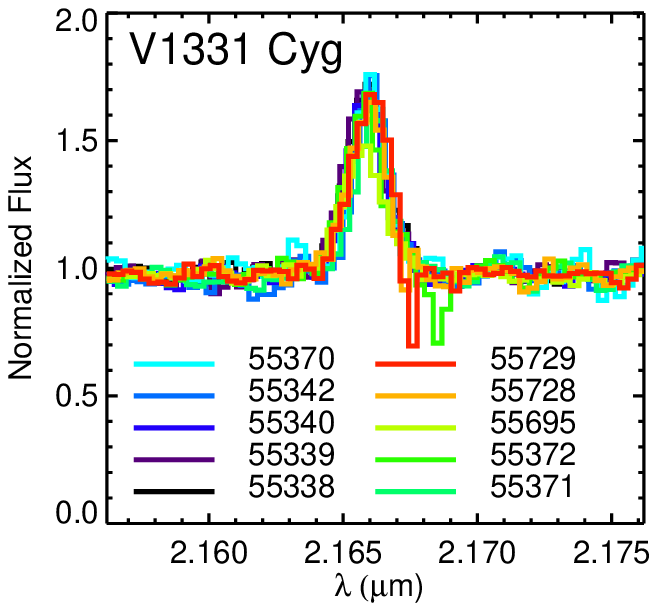}
\caption{Spectra of V1331 Cyg, the heavily-veiled star in our
  sample whose spectral type is uncertain.   Observed epochs
  are color-coded by MJD.
\label{fig:v1331}}
\end{figure}

As one of the fainter objects in our sample, the spectra for V1331 Cyg
have somewhat lower signal-to-noise.  However, the Br$\gamma$ emission
from this target is strong and easily detectable above the noise level
(Figure \ref{fig:v1331}).  The EWs vary from $-9 \pm 1$ to $-14 \pm 1$
\AA ~across the $>1$ year of observations (Table \ref{tab:sample}).   Previous
measurements of Br$\gamma$ emission estimated EWs of -15.3 \AA ~in
1994 \citep{NCT96}, and -14.4 \AA ~in 2009 \citep{EHS14}.  While these
previous measurements, and most of our spectra, show single-peaked
emission line profiles, red absorption features appear strongly in one
epoch and tentatively in another epoch (Figure \ref{fig:v1331}).  Such
line profiles resemble inverse P Cygni profiles, and may indicate that
we are seeing red absorption from infalling matter along the line of sight.

\section{DISCUSSION}
\label{sec:disc}
\subsection{Br$\gamma$ Emission and Variability Across the Sample}
Our sample spans a broad range of spectral type, from solar-type T
Tauri stars to $\ga 10$ M$_{\odot}$ stars.  The sample also includes
heavily-veiled objects, which are presumably surrounded by massive,
active accretion disks.  The extreme end of the sample comprises FU
Ori sources, where the central stars are completely invisible underneath
the hot, active accretion disks.  Here we examine measured Br$\gamma$
EWs and line profiles, and their variability, across this diverse
sample.

We detect Br$\gamma$ emission around 4/5 of the T Tauri stars in our
sample: only V1002 Sco shows no emission.  The Br$\gamma$ emission
from these sources has a typical EW with absolute value smaller 
than about  $5$ \AA ~\citep[consistent with previous Br$\gamma$
spectroscopy; e.g.,][]{MHC98}.  Variability,
when detected, is modest, at the level of $\sim 2$ \AA.

Among the Herbig Ae sources in our sample, 6/6 objects have detected
Br$\gamma$ emission.  Typical EWs and variability amplitudes are
similar to those seen in the T Tauri sub-sample.  We do, however, note
the broad, double-peaked line profiles seen in the Herbig Ae sources
HD 141569 and 51 Oph, perhaps suggesting a different morphology of the
emission in these objects.

Br$\gamma$ emission is detected toward all Herbig Be sources except
two: photospheric absorption is 
seen in LkH$\alpha$ 169 and no emission or absorption is detected
toward V361 Cep.  In contrast to the T Tauri and Herbig Ae stars,
Br$\gamma$ emission from the Herbig Be sample tends
to be stronger and more variable.  For example, a number of objects
have Br$\gamma$ EWs of -10 to -20 \AA, and variability up to 5 \AA 
~is observed in some targets.  The strongest Br$\gamma$ emission is
observed from the early-type objects in our sample, consistent with
previous results \citep[e.g.,][]{GARCIALOPEZ+06}.

The most massive star in the sample, Herbig Be source V645
Cyg, exhibits a line profile suggestive of outflow (Figure
\ref{fig:hbes}).  In particular, the blue absorption feature seen in
most epochs indicates
outflowing matter in the observed line-of-sight that is absorbing
background emission.  MWC 297, the second-most-massive star in the
sample, does not show such a line profile.  However viewing geometry
may lead to outflows outside the line-of-sight, and hence a lack of
absorption features.  Indeed, spatially resolved
spectroscopy of MWC 297 has suggested that Br$\gamma$ gas arises in
a magnetocentrifugally-launched wind on scales of $\ga 1$ AU
\citep{MALBET+07,WEIGELT+11}.  Thus, the Br$\gamma$ emission we
observe from the most massive Herbig Be stars in our sample 
may trace outflows more
commonly than the emission seen in less massive objects.

V1331 Cyg is an unusual object compared to others in the sample.  It
is heavily veiled, and hence its spectral type is unknown.  It also
exhibits strong CO overtone emission.  These properties suggest a
massive and/or hot accretion disk around the central star.  As in
previous work, we find strong Br$\gamma$ emission from V1331
Cyg, consistent with the hypothesis of an active accretion disk around
the source.  We also see red absorption features in some observed
epochs, suggesting infalling matter along the line-of-sight.  The
observed line-profile and EW variability indicates that this infall
may be sporadic.

Finally, we included two FU Ori stars in our sample: V1057 Cyg and
V1515 Cyg.  These objects are characterized by spectra that resemble
stellar photospheres at several thousand degrees
\citep[e.g.,][]{HK96}.  At such temperatures, one does not expect
significant Br$\gamma$ absorption or emission.  That is essentially
what we see in the FSPEC data.  However, both objects show hints of
features, with possible weak absorption with redward emission in V1057
Cyg, and a weak emission feature in V1515 Cyg
(Figure \ref{fig:fuors}).

\subsection{Br$\gamma$ Emission and Variability as a Tracer of
  Magnetospheric Accretion}
As discussed in Section \ref{sec:intro}, Br$\gamma$ is empirically correlated
with mass accretion rate onto young stars \citep[][]{MHC98,MENDIGUTIA+11}.
While this is also true for optical emission lines like H$\alpha$ 
\citep[e.g.,][]{GULLBRING+98}, comparison of optical and near-IR line
profiles suggests that the former are more likely to trace wind
emission, and the latter more likely to trace infalling matter
\citep[e.g.,][]{NCT96,FE01,MCH01}.  

Imaging observations have suggested that Br$\gamma$ may still trace
outflowing gas near the magnetospheric scale
\citep[e.g.,][]{EISNER+10,EHS14},  or in some cases on considerably
larger scales \citep[e.g.,][]{MALBET+07,BBM10}.  However, in nearly all objects
where Br$\gamma$ emission has been spatially resolved, some of the
emission traces very compact scales consistent with accretion flows
\citep[e.g.,][]{EHS14}.  Furthermore, spatially extended emission is
generally found to constitute only a small fraction of the total
Br$\gamma$ line flux \citep[e.g.,][]{BBM10}.  Thus, despite evidence
that {\it some} Br$\gamma$ emission originates in outflows, it appears
that Br$\gamma$ is still a better tracer of accretion flows than
optical lines like H$\alpha$.

Variability in Br$\gamma$ emission can occur for a variety of
reasons.  Since hydrogen line emission 
traces infalling material, when accretion rates are higher more
hydrogen gas is excited and produces more Br$\gamma$ emission.
Alternatively, a steady global disk accretion rate may interact with a
rotating, magnetic star and produce time-variable hotspots or
accretion streams \citep[e.g.,][]{ROMANOVA+04,RKL08}. Indeed, models
of disk/magnetosphere interactions show that Br$\gamma$ line profiles
and EWs can vary on stellar rotation timescales \citep[e.g.,][]{KRH08}.
However variations on shorter timescales are also possible for
unstable accretion, where multiple, thin, variable accretion flows
impact the star \citep{KR13}.

Magnetically-mediated accretion models thus predict that Br$\gamma$
emission should vary on timescales comparable to or shorter than
stellar rotation periods.  Measured rotation periods for young stars
are typically $\sim 1$--10 days \citep[e.g.,][]{HERBST+07}.  While we
argued in Section \ref{sec:results} that many objects show variations
on night-to-night timescales, a rigorous comparison with models
requires quantification of the observed variability periods.

Quantitative constraints are difficult with the limited time-sampling
of our data.  We therefore restrict our quantitative analysis to those
objects with the most data.  We select targets with
significant variability in their Br$\gamma$ emission, for which we
obtained $> 5$ epochs of data.  We then determine
frequency distributions of measured EWs using periodograms
\citep{LOMB76,SCARGLE82}.  

\begin{figure*}
\plotone{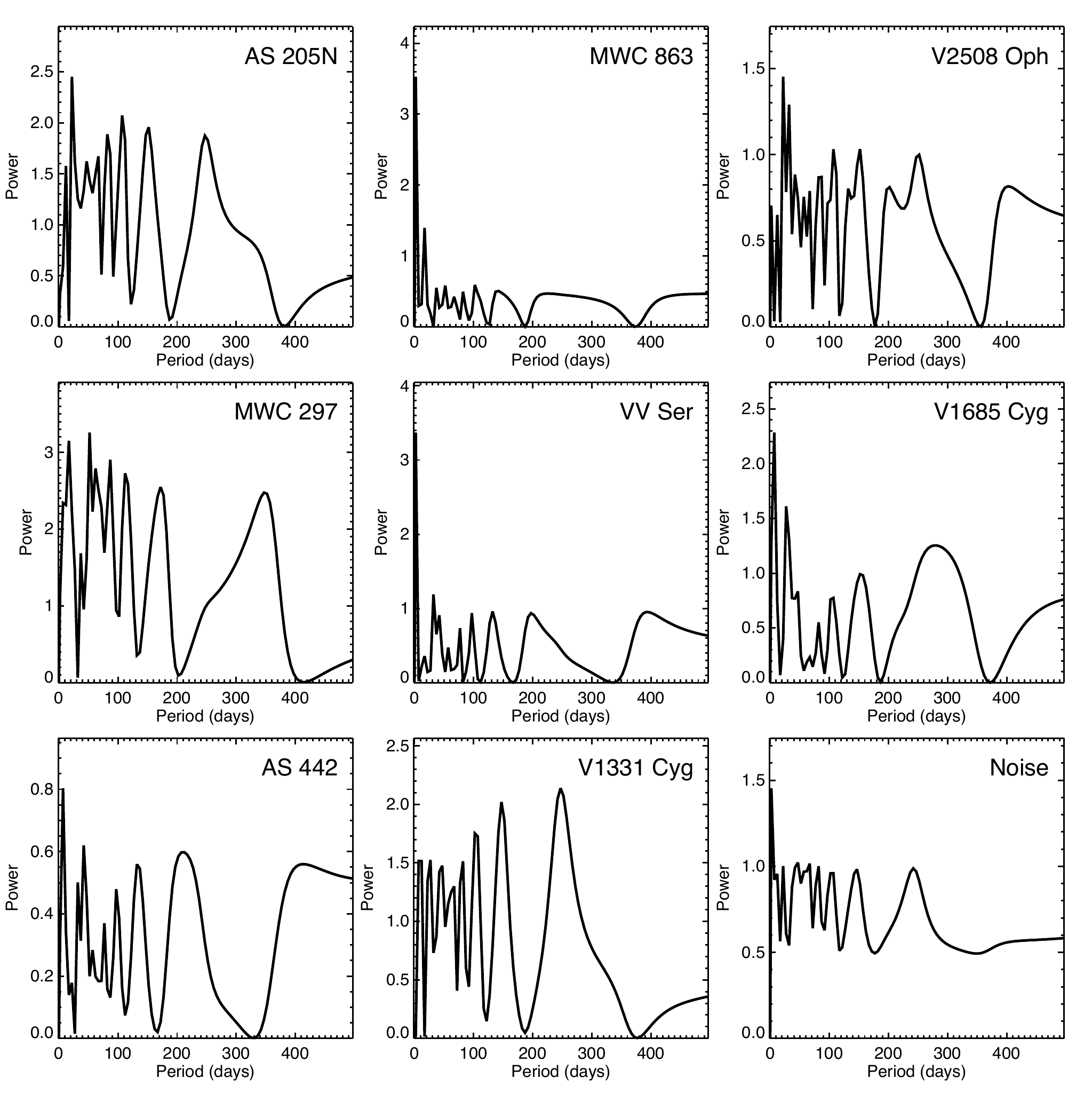}
\caption{Frequency distributions of measured EWs over multiple epochs.
   These distributions were calculated using periodograms
   \citep{LOMB76,SCARGLE82}, and were determined for targets that
  show significant variabiltiy in Br$\gamma$ EW over at least 5
  observed epochs (Table \ref{tab:sample}).  The bottom right
  panel shows the average of 1000 periodograms, each calculated for a
  single realization of Gaussian noise sampled at the same epochs as
  the data for V1331 Cyg.
\label{fig:periodograms}}
\end{figure*}

Figure \ref{fig:periodograms} shows the calculated periodograms for 8
of our sample objects.  In general we see more power on the shortest
periods ($<5$ days) than on other timescales.  This is consistent with
the observation that variability appears to occur on night-to-night
timescales in many sources, and is compatible with variability
associated with magnetospheric accretion processes
\citep[e.g.,][]{KR13}.  

\epsscale{1.0}

However the time-sampling of the observations
alone can lead to power at short periods in the periodograms.  To test
this, we calculated the periodograms for 1000 realizations of
Gaussian noise, with the same time-sampling as our observations.  The
average of these 1000 periodograms is included in Figure
\ref{fig:periodograms}.  The periodograms for several targets, most
notably V1685 Cyg and AS 442, resemble the one computed for Gaussian
noise.  For these sources, the variability may be random, with no
preferred timescale.

The periodograms for other objects appear distinct from the one
determined for Gaussian noise, and suggest variability on longer
timescales.  V1331 Cyg
shows a peak in its periodogram on a timescale $>200$ days.  MWC 297
also shows power on timescales $>200$ days at approximately the same
level as the power seen on shorter variability timescales.  We noted in
Section \ref{sec:results} that our observed EWs for MWC 297 differ
significantly from previously published measurements from several
years earlier.  These long-term variations are probably not associated 
with star/disk interactions.  However secular changes in the accretion
rate may be associated with this longer-timescale variability.

Magnetospheric accretion models also
make predictions about Br$\gamma$ line profiles and their
variability \citep{KRH08,KR13}.  
Synthetic Br$\gamma$ emission lines show redshifted
absorption components.  Redshifted
absorption may appear and disappear as accretion flows rotate in and
out of the observed line-of-sight, or may persist for accretion in the
unstable regime where multiple, thin, variable accretion flows
impact the star.  These absorption components are 
weaker, or absent, in synthetic spectra of H$\alpha$
and other optical lines.  The Br$\gamma$ absorption features are
stronger, and variability amplitudes of line profiles higher,
because Br$\gamma$ traces matter near the base of the accretion stream
at high velocities, while H$\alpha$ traces Stark-broadened emission
from a much larger area.  

While the redshifted absorption components and line profile changes
predicted by models would be easily detectable in our FSPEC data, we
observe a paucity of either.
We see little redshifted absorption in the Br$\gamma$ lines for our
sample.  V1331 Cyg shows strong redshifted absorption---in only some
epochs---and V2508 Oph shows weak redshifted absorption.  This
represents a detection rate of $\sim 10\%$ for the sources in our
sample exhibiting Br$\gamma$ emission.
For comparison, \citet{FE01} saw redshifted Br$\gamma$
absorption components around 20\% of their sample.   Given the lower
spectral resolution and signal-to-noise in our data compared to those
of \citet{FE01}, and the fairly weak absorption features seen in some
objects by those authors,  our detection rate is probably consistent
with theirs.

Studies of optical emission lines, including those in the Balmer
series, often claim much higher rates of redshifted absorption. For
example, a study of multiple optical emission lines found redshited
H$\delta$ absorption in $>50 \%$ of observed objects
\citep[e.g.,][]{EDWARDS+94}.  
Red absorption features are seen to persist across multiple epochs in
some sources \citep[e.g.,][]{EDWARDS+94} or to appear and disappear in other
objects \citep[e.g.,][]{BERTOUT+77}.  Simultaneous monitoring of
optical and near-IR lines shows that the amplitude of
variability in optical lines is similar to, or marginally larger than
the variability amplitude of near-IR lines like
Br$\gamma$ \citep[e.g.,][]{MENDIGUTIA+13}.

Magnetospheric accretion models thus make several predictions
\citep[e.g.,][]{KR13} that can
be compared with our observations.  Expected variability timescales
are on the order of stellar rotation periods, and our inferred variability
timescales are roughly consistent with this expectation.
However the predicted red absorption features, and their
time-variability, is not consistent with our data.  Furthermore, the 
greater occurence rate of such spectral features in optical lines
compared to Br$\gamma$ is also inconsistent with models.
Thus it appears that
the models must be modified to account for both the optical and
near-IR emission lines, or that a
different physical process may be responsible for the Br$\gamma$
emission variability seen from young stars.

\subsection{Time-Variable Extinction or Continuum Level}
Extinction can complicate the picture outlined above by obscuring the
central stars or accretion flows, 
especially for nearly edge-on disks where the
line of sight to the central star intersects part of the dusty inner disk.
This obscuration may actually increase for higher accretion rates,
since increased viscous dissipation heats the inner disk and causes a
hydrostatic increase in the disk scale height \citep[e.g., ][]{STONE+14}.
In this scenario, Br$\gamma$ emission from infalling matter may be
observed to decrease even as $\dot{M}$ increases, because of the
increased extinction.

Extinction may also lead to a situation where observed variability in
Br$\gamma$ emission is dominated by variability in outflows rather
than accretion streams.  Since the inner disk scale height is large
compared to the size of the central star, even magnetospheric
accretion flows onto polar regions of the star can be obscured by a
nearly edge-on disk.  While the mass accretion rate is
typically higher than the mass outflow rate \citep[by about a factor
of 10; e.g.,][]{HEG95}, if the accretion flow is extincted
while the outflow is not, then any observed
variability may be dominated by processes in the outflow.  Since
extinction of the inner accretion stream could increase with
$\dot{M}$, the relative contributions of infall and outflow to the
observed Br$\gamma$ line profile might also vary with $\dot{M}$.

Finally, extinction may obscure the central star while leaving more of
the accretion flow visible.  In this case, the continuum level might
decrease while the Br$\gamma$ line flux (and line profile) 
remained approximately constant.  For the majority of our
sample, the $K$-band continuum flux is dominated by emission from
circumstellar matter \citep[e.g.,][]{EHS14}, and thus any variable
obscuration of the central star is expected to be a minor effect.

However, changes in continuum level due to variations in
the structure of the inner disk \citep[e.g.,][]{FLAHERTY+13,CODY+14},
may also be important for interpreting our observations.  Continuum
changes would cause variability in the Br$\gamma$ line-to-continuum
ratios, and hence in the measured EWs.

Some studies have found evidence for photometric and spectroscopic
variability due to time-variable obscuration by a warped inner disk on
stellar rotation (or equivalently, corotation) timescales
\citep[e.g.,][]{BOUVIER+07}.  In general, however, one might
expect variability on the dynamical timescales
corresponding to the inner edge of the {\it dust} disk, which is
typically larger (by factors up to a few) than the corotation radius
\citep[e.g.,][]{EISNER+05}.  It is this dusty inner disk that
dominates the continuum emission and the potential extinction.

With the time-sampling of our data, we are not sensitive to
the difference in 
period expected for variability near the inner edge
of the dust disk and variability linked to the magnetospheric scale.
However, the lack of line profile variability seen in many objects is
consistent with time-variable extinction or continuum level, 
but not with models of
variable, magnetically-mediated accretion.

\section{Conclusions}
We presented spectra of a sample of 25 young stars, covering the
Br$\gamma$ feature at a resolution of $\approx 3500$.  For 10 of these
targets, the data presented here are the first observations of the
Br$\gamma$ spectral region, and emission was detected from 8 for the
first time.  For the other targets in our sample, which had been
previously observed, we presented multiple epochs of new data, which
we used to constrain the time-variability of Br$\gamma$ emission from
our sample.

For most targets, observations were taken over a span of
$>$1 year, sampling time-spacings of roughly one day, one month,
and one year.  We detected Br$\gamma$ emission from 21 sources, and
saw significant variability in 10 of them (approximately 50\%).  This
variability is constrained quantitatively through measurements of EW.
In some objects, we also observed changes in line profiles with time.

The observed variability seems to occur over a range of timescales.
Night-to-night variations were detected in some objects, and a
periodogram analysis suggested that some variability 
occurs on such timescales.  These timescales are similar to stellar
rotation periods, and are compatible with those
expected for rotationally-modulated accretion onto the central
stars.  However the line profiles---which typically lack redshifted
absorption across all observed epochs---are inconsistent with
rotationally-modulated accretion models.  Furthermore, some targets in
our sample show evidence for substantial variations in Br$\gamma$ EW
and line profiles on timescales of months or years, much longer than
stellar rotation periods.

An alternative explanation for observed Br$\gamma$ variability is
time-variable extinction or changes in the continuum flux level.  
Such variability might occur on inner disk
dynamical timescales, which are typically days to weeks, or on longer
timescales if there are orbiting clouds of dust at larger radii.
While continuum changes or time-variable extinction could lead to
variability in Br$\gamma$ EW, line profiles would remain approximately
constant.  This scenario appears consistent with the observed
variability for many of our targets. 

\noindent{\bf Acknowledgments.}
JAE gratefully acknowledges support from an Alfred P. Sloan Research
Fellowship.  Support for this work, largely for students affiliated
with the project, comes from the National Science
Foundation under Award No. AST-0847170, a PAARE Grant for the
California-Arizona Minority Partnership for Astronomy Research and
Education (CAMPARE).   Any opinions, findings, and conclusions or
recommendations expressed in this material are those of the author(s)
and do not necessarily reflect the views of the National Science
Foundation.

\bibliographystyle{apj} 
\bibliography{jae_ref}

\LongTables

\begin{deluxetable*}{lccccc|ccc|c}
\tabletypesize{\scriptsize}
\tablewidth{0pt}
\tablecaption{Observed Targets, Epochs, and Measured EWs
\label{tab:sample}}
\tablehead{\colhead{Source} & \colhead{$\alpha$} & \colhead{$\delta$}
  & \colhead{MJD} & \colhead{$T_{\rm
      int}$} & \colhead{Airmass} & \colhead{Calibrators}
  &\colhead{$T_{\rm int}$} & \colhead{Airmass}  & \colhead{Br$\gamma$ EW} \\
 & (J2000) & (J2000) & & (s) & &  & (s) & & (\AA) }
\startdata
\multicolumn{10}{c}{\bf T Tauri Stars} \\
\hline \hline
AS 205 N & 16 11 31.40 & -18 38 24.5 & 55338 & 450 & 1.6 & HD
144253 & 96 & 1.6 & $-6 \pm 1$ \\
& & & 55339 & 900 & 1.6 & HD 144253 & 240 & 1.7 & $-7 \pm 1$ \\ 
& & & 55340 & 300 & 1.6 & HD 144253 & 96 & 1.6 & $-5 \pm 1$ \\
& & & 55341 & 300 & 1.6 & HD 144253 & 120 & 1.7 & $-4 \pm 1$ \\
& & & 55342 & 300 & 1.7 & HD 144253 & 192 & 1.7 & $-5 \pm 1$ \\ 
& & & 55373 & 300 & 1.6  & HD 144253 & 160 & 1.7 & $-3 \pm 1$ \\
& & & 55694 & 150 & 1.6 & HD 144253 & 120 & 1.6 & $-4 \pm 1$ \\
& & & 55695 & 300 & 1.7 & HD 144253 & 120 & 1.8 & $-2 \pm 1$ \\
& & & 55728 & 250 & 1.6 & HD 144253 & 125 & 1.7 & $-5 \pm 1$ \\
& & & 55729 & 250 & 1.6 & HD 144253 & 125 & 1.7 & $-4 \pm 1$ \\
\hline
V1002 Sco & 16 12 40.51 & -18 59 28.1 & 55339 &  900 & 1.6 & HD
144253 & 240 & 1.7 
& $-1 \pm 1$ \\
& & & 55340 & 450 & 1.6 & HD 144253 & 96 & 1.6 & $0 \pm 1$ \\
& & & 55341 & 600 & 1.6 & HD 144253 & 120 & 1.7 & $0 \pm 1$ \\ 
& & & 55342 &  450 & 1.6 & HD 144253 & 192 & 1.7 & $0 \pm 1$ \\
& & & 55694 & 300 & 1.6 & HD 144253 & 120 & 1.6 & $0 \pm 1$ \\
& & & 55695 & 300 & 1.7 & HD 144253 & 120 & 1.8 & $0 \pm 1$ \\
& & & 55728 & 375 & 1.6 & HD 144253 & 125 & 1.7 & $-1 \pm 2$ \\
& & & 55729 & 375 & 1.6 & HD 144253 & 125 & 1.7 & $0 \pm 1$ \\
\hline
V2508 Oph & 16 48 45.63 & -14 16 40.0 & 55338 & 450 & 1.5 & HD
144253 & 96 & 1.6 & $-4 \pm 1$ \\
& & & 55339 & 900 & 1.4 & HD 144253 & 240 & 1.7 & $-3 \pm 1$ \\
& & & 55340 &  450 & 1.5 & HD 144253 & 96 & 1.6 & $-2 \pm 1$ \\
& & & 55341 &  300 & 1.5 & HD 144253 & 120 & 1.7 & $-3 \pm 1$ \\
& & & 55342 &  450 & 1.5 & HD 144253 & 192 & 1.7 & $-2 \pm 1$ \\
& & & 55694 &  450 & 1.5 & HD 144253 & 120 & 1.6 & $-3 \pm 1$ \\
& & & 55695 &  300 & 1.6 & HD 144253 & 120 & 1.8 & $0 \pm 1$ \\
& & & 55728 & 250 & 1.4 & HD 144253 & 125 & 1.7  & $-3 \pm 1$ \\
& & & 55729 &  250 & 1.4 & HD 144253 & 125 & 1.7 & $-2 \pm 1$ \\
\hline
AS 209 & 16 49 15.30 & -14 22 08.6 & 55338 & 450 & 1.5 & HD 151528
& 120 & 1.5 & $-3 \pm 1$ \\
& & & 55339 & 600 & 1.5 & HD 151528 & 300 & 1.5 & $-3 \pm 1$ \\
& & & 55340 &  300 & 1.5 & HD 151528 & 240 & 1.5 & $-3 \pm 1$ \\
& & & 55341 &  300 & 1.5 & HD 151528 & 150 & 1.5 & $-3 \pm 1$ \\
& & & 55694 &  300 & 1.5 & HD 151528 & 240 & 1.5 & $-2 \pm 1$ \\
& & & 55695 &  300 & 1.6 & HD 151528 & 150 & 1.6 & $-3 \pm 1$ \\
& & & 55728 &  250 & 1.5 & HD 151528 & 125 & 1.5 & $-3 \pm 1$ \\
& & & 55729 & 375 & 1.5 & HD 151528 & 125 & 1.5 & $-3 \pm 1$ \\
\hline
V521 Cyg & 20 58 23.81 & 43 53 11.4 & 06/24/10 & 300 & 1.1 & HD 192164
& 60 & 1.0 & $-2 \pm 1$ \\
& & & 55373 & 200 & 1.0 & HD 192164 & 80 & 1.1 & $-3 \pm 2$ \\
\hline  \hline
\multicolumn{9}{c}{\bf Herbig Ae Stars} \\
\hline \hline
HD 141569 & 15 49 57.75 & -03 55 16.3 & 55338 & 600 & 1.3 & HD
135204 & 96 & 1.2 & $6 \pm 1$ \\
 & & & 55339 & 840 & 1.3 & HD 135204 & 120 & 1.2 & $6 \pm 1$ \\
 & & & 55342 & 240 & 1.3 & HD 135204 & 120 & 1.2 & $7 \pm 1$ \\
 & & & 55694 & 420 & 1.3 & HD 135204 & 180 & 1.2 & $7 \pm 1$ \\
 & & & 55695 & 300 & 1.3 & HD 135204 & 150 & 1.2 & $5 \pm 1$ \\
 & & & 55728 & 250 & 1.2 & HD 135204 & 125 & 1.2 & $6 \pm 1$ \\
 & & & 55729 & 250 & 1.2 & HD 135204 & 125 & 1.2 & $6 \pm 1$ \\
\hline
MWC 863 & 16 40 17.92 & -23 53 45.2 &  55338 & 120 & 1.8 & HD
154088 & 84 & 2.0 & $-3 \pm 1$ \\
& & & 55341 & 120 & 1.8 & HD 154088 & 84 & 2.0 & $-5 \pm 1$ \\
& & & 55694 &  240 & 1.8 & HD 154088 & 90 & 2.0 & $-3 \pm 1$ \\
& & & 55728 &  125 & 1.8 & HD 154088 & 125 & 2.0 & $-2 \pm 1$ \\
& & & 55729 & 250 & 1.9 & HD 154088 & 125 & 2.0  & $-4 \pm 1$ \\
\hline
51 Oph & 17 31 24.95 & -23 57 45.5 & 55338 & 60 & 1.8 & HD 157172
&120 & 1.6 & $4 \pm 1$ \\
& & & 55341 &  120 & 1.8 & HD 157172 & 150 & 1.6 & $2 \pm 1$ \\
& & & 55694 &  120 & 1.8 & HD 157172 & 270 & 1.6 & $4 \pm 1$ \\
& & & 55695 &  150 & 1.8 & HD 157172 & 300 & 1.6 & $4 \pm 1$ \\
& & & 55728 & 125 & 1.9 & HD 157172 & 250 & 1.6  & $3 \pm 1$ \\
& & & 55729 & 250 & 1.8 & HD 157172 & 250 & 1.6  & $4 \pm 1$ \\
\hline
MWC 275 & 17 56 21.29 & -21 57 21.9 & 55338 & 216 & 1.7 & HD 157172
& 120 & 1.6 & $-6 \pm 1$ \\
& & & 55341 &  120 & 1.7 & HD 157172 & 150 & 1.6 & $-7 \pm 1$ \\
& & & 55694 &  120 & 1.7 & HD 157172 & 270 & 1.6 & $-5 \pm 1$ \\
& & & 55695 &  150 & 1.8 & HD 157172 & 300 & 1.6 & $-6 \pm 1$ \\
& & & 55728 & 125 & 1.7 & HD 157172 & 300 & 1.6 & $-6 \pm 1$ \\
& & & 55729 & 125 & 1.7 & HD 157172 & 250 & 1.6 & $-5 \pm 3$ \\
\hline
VV Ser & 18 28 47.90 & 00 08 40.0 & 55338 & 120 & 1.2 & HD 174719 &
120 & 1.2 & $-6 \pm 1$ \\
& & & 55339 & 600 & 1.2 & HD 174719 & 300 & 1.2 & $-7 \pm 1$ \\
& & & 55340 &  120 & 1.2 & HD 174719 & 120 & 1.2 & $-5 \pm 1$ \\
& & & 55341 &  120 & 1.2 & HD 174719 & 120 & 1.2 & $-9\pm 1$ \\
& & & 55371 &  360 & 1.2 & HD 174719 & 120 & 1.1 & $-5 \pm 1$ \\
& & & 55694 &  270 & 1.2 & HD 174719 & 120 & 1.2 & $-7 \pm 1$ \\
& & & 55695 &  300 & 1.2 & HD 174719 & 150 & 1.3 & $-8 \pm 1$ \\
& & & 55728 &  250 & 1.2 & HD 174719 & 250 & 1.1 & $-7 \pm 1$ \\
& & & 55729 &  250 & 1.2 & HD 174719 & 125 & 1.1 & $-7 \pm 1$ \\
\hline
V2020 Cyg & 20 48 20.35 & 43 39 48.2 & 55338 & 120 & 1.1 & HD
204814 & 60 & 1.0 & $-5 \pm 1$ \\
& & & 55370 & 300 & 1.0 & HD 192164 & 120 & 1.0 & $-7 \pm 1$ \\
& & & 55371 & 300 & 1.0 & HD 192164 & 120 & 1.0 & $-5 \pm 1$ \\
& & & 55373 & 300 & 1.0 & HD 192164 & 80 & 1.1 & $-5 \pm 1$ \\
\hline  \hline
\multicolumn{9}{c}{\bf Herbig Be Stars} \\
\hline \hline
EU Ser & 18 19 09.38 & -13 50 41.1 & 55371 & 450 & 1.5 & HD 174719
& 120 & 1.1 & $-3 \pm 1$ \\
& & & 55373 & 200 & 1.4 & HD 174719 & 80 & 1.1 & $-3 \pm 2$ \\
\hline
MWC 297 & 18 27 39.53 & -03 49 52.1 & 55338 & 36 & 1.2 & HD 174719
& 120 & 1.2 & $-26 \pm 1$ \\
& & & 55339 & 144 & 1.3 & HD 174719 & 300 & 1.2 & $-26 \pm 1$ \\
& & & 55340 &  36 & 1.2 & HD 174719 & 120 & 1.2 & $-26 \pm 1$ \\
& & & 55341 & 36 & 1.2 & HD 174719 & 120 & 1.2 & $-28 \pm 1$ \\
& & & 55371 &  72 & 1.2 & HD 174719 & 120 & 1.1 & $-22 \pm 1$ \\
& & & 55694 &  126 & 1.2 & HD 174719 & 120 & 1.2 & $-24 \pm 1$ \\
& & & 55695 &  78 & 1.3 & HD 174719 & 150 & 1.3 & $-26 \pm 1$ \\
& & & 55728 & 80 & 1.2 & HD 174719 & 250 & 1.1 & $-24 \pm 1$ \\
& & & 55729 & 36 & 1.2 & HD 174719 & 125 & 1.1 & $-23 \pm 1$ \\
\hline
V1685 Cyg & 20 20 28.24 & 41 21 51.6 & 55338 & 96 & 1.1 & HD 204814
& 60 & 1.0 & $-12 \pm 1$ \\
& & & 55339 & 240 & 1.0 & HD 204814 & 240 & 1.1 & $-12 \pm 1$ \\
& & & 55340 &  120 & 1.0 & HD 204814 & 96 & 1.1 & $-13 \pm 1$ \\
& & & 55341 &  120 & 1.1 & HD 204814 & 96 & 1.2 & $-14 \pm 1$ \\
& & & 55342 &  120 & 1.0 & HD 192164 & 60 & 1.0 & $-9 \pm 1$ \\
& & & 55370 & 120 & 1.0 & HD 192164 & 120 & 1.0 & $-13 \pm 1$ \\
& & & 55371 & 120 & 1.0 & HD 192164 & 120 & 1.0 & $-11 \pm 1$ \\
& & & 55372 & 120 & 1.0 & HD 192164 & 60 & 1.0 & $-11 \pm 1$ \\
& & & 55373 & 80 & 1.1 & HD 192164 & 80 & 1.1 & $-11 \pm 1$ \\
& & & 55695 &  150 & 1.1 & HD 192164 & 90 & 1.2 & $-10 \pm 1$ \\
& & & 55728 &  125 & 1.0 & HD 192164 & 100 & 1.0 & $-10 \pm 1$ \\
\hline
V1972 Cyg & 20 23 03.61 & 39 29 50.1 & 55338 & 60 & 1.0 & HD 204814
& 60 & 1.0 & $-17 \pm 1$ \\
& & & 55339 & 240 & 1.0 & HD 204814 & 240 & 1.1 & $-16 \pm 1$ \\
& & & 55340 &  60 & 1.0 & HD 204814 & 96 & 1.1 & $-23 \pm 1$ \\
& & & 55341 & 60 & 1.0  & HD 204814 & 96 & 1.2  & $-19 \pm 1$ \\
\hline
AS 442 & 20 47  37.47 & 43 47 24.9 & 55338 & 240 & 1.1 & HD 204814
& 60 & 1.0 & $-7 \pm 1$ \\
& & & 55339 & 300 & 1.0 & HD 204814 & 240 & 1.1 & $-3 \pm 1$ \\
& & & 55340 &  240 & 1.1 & HD 204814 & 96 & 1.1 & $-3 \pm 1$ \\
& & & 55341 &  240 & 1.1 & HD 204814 & 96 & 1.2 & $-5 \pm 1$ \\
& & & 55342 &  240 & 1.0 & HD 192164 & 60 & 1.0 & $-5 \pm 1$ \\
& & & 55370 & 120 & 1.0 & HD 192164 & 120 & 1.0 & $-6 \pm 1$ \\
& & & 55371 & 120 & 1.0 & HD 192164 & 120 & 1.0 & $-4 \pm 1$ \\
& & & 55373 & 160 & 1.0 & HD 192164 & 80 & 1.1 & $-6 \pm 2$ \\
& & & 55695 &  150 & 1.2 & HD 192164 & 90 & 1.2 & $-4 \pm 1$ \\
& & & 55728 & 375 & 1.0 & HD 192164 & 100 & 1.0  & $-4 \pm 1$ \\
\hline
V645 Cyg & 21 39 58.24 & 50 14 21.2 & 55338 & 120 & 1.1 & HD 192164
& 60 & 1.0 & $-2 \pm 1$ \\
& & & 55370 & 150 & 1.1 & HD 192164 & 120 & 1.0 & $-2 \pm 1$ \\
& & & 55371 & 150 & 1.1 & HD 192164 & 120 & 1.0 & $-1 \pm 1$ \\
& & & 55373 & 400 & 1.1 & HD 192614 & 80 & 1.1 & $-2 \pm 1$ \\
\hline
V2019 Cyg & 20 48 04.79 & 43 47 25.8 & 55338 & 270 & 1.1 & HD
204814 & 60 & 1.0 & $-17 \pm 1$ \\
& & & 55339 & 480 & 1.0 & HD 204814 & 240 & 1.1 & $-21 \pm 1$ \\
& & & 55340 & 450 & 1.0 & HD 204814 & 96 & 1.1  & $-21 \pm 1$ \\
& & & 55341 &  300 & 1.1  & HD 204814 & 96 & 1.2 & $-18 \pm 1$ \\
\hline
LkH$\alpha$ 169 & 20 52 07.65 & 44 03 44.4 & 55338 & 150 & 1.1 & HD
190113 & 36 & 1.0 & $14 \pm 2$ \\
& & & 55341 & 240 & 1.0 & HD 204814 & 90 & 1.1 & $11 \pm 2$ \\
& & & 55370 & 450 & 1.0 & HD 192164 & 120 & 1.0 & $11 \pm 2$ \\
\hline
V380 Cep & 21 01 36.92 & 68 09 47.8 & 55373 & 300 & 1.2 & HD 199476
& 80 & 1.4 & $-10 \pm 3$ \\
\hline
V361 Cep & 21 42 50.18 & 66 06 35.1 & 55373 & 400 & 1.2 & HD 199476
& 80 & 1.4 & $1 \pm 1$ \\
\hline
MWC 1080 & 23 17 25.57 & 60 50 43.3 & 55373 & 200 & 1.2 & HD 215500
& 80 & 1.2 & $-7 \pm 2$ \\
\hline  \hline
\multicolumn{9}{c}{\bf FU Ori Objects} \\
\hline \hline
V1057 Cyg & 20 58 53.73 & 44 15 28.5 & 55338 & 120 & 1.2 & HD
192164 & 120 & 1.1 & $0 \pm 1$ \\
& & & 55339 & 600 & 1.2 & HD 192164 & 120 & 1.1 & $0 \pm 1$ \\
& & & 55340 &  120 & 1.1 & HD 192164 & 60 & 1.0 & $-2 \pm 2$ \\
& & & 55341 &  120 & 1.2 & HD 192164 & 60 & 1.1 & $0 \pm 1$ \\
& & & 55342 &  240 & 1.1 & HD 192164 & 60 & 1.0 & $2 \pm 2$ \\
& & & 55370 & 150 & 1.0 & HD 192164 & 120 & 1.0 & $0 \pm 1$ \\
& & & 55371 & 150 & 1.1 & HD 192164 & 120 & 1.0 & $1 \pm 1$ \\
& & & 55694 & 270 & 1.1 & HD 192164 & 60 & 1.1  & $0 \pm 1$ \\
& & & 55695 & 150 & 1.3 & HD 192164 & 90 & 1.2  & $0 \pm 1$ \\
\hline
V1515 Cyg & 20 23 48.02 & 42 12 25.8 &  55338 & 240 & 1.1 & HD
190113 & 156 & 1.1 & $-1 \pm 1$ \\
& & & 55341 & 360 & 1.1 & HD 190113 & 276 & 1.0 & $-1 \pm 1$ \\
& & & 55342 &  450 & 1.1 & HD 190113 & 300 & 1.1 & $-2 \pm 1$ \\
& & & 55370 & 600 & 1.1 & HD 190113 & 270 & 1.1 & $-2 \pm 1$ \\
& & & 55371 & 300 & 1.0 & HD 190113 & 120 & 1.0 & $-1 \pm 1$ \\
& & & 55694 &  300 & 1.1 & HD 190113 & 270 & 1.1 & $0 \pm 1$ \\
& & & 55695 &  300 & 1.2 & HD 190113 & 150 & 1.2 & $0 \pm 1$ \\
\hline  \hline
\multicolumn{9}{c}{\bf Heavily Veiled Source} \\
\hline \hline
V1331 Cyg & 21 01 09.21 & 50 21 44.8 &  55338 & 450 & 1.1 & HD
192164 & 120 & 1.1 & $-14 \pm 1$ \\
& & & 55339 & 900 & 1.1 & HD 192164 & 120 & 1.1 & $-14 \pm 1$ \\
& & & 55340 &  450 & 1.1 & HD 192164 & 60 & 1.0 & $-13 \pm 1$ \\
& & & 55342 &  300 & 1.1 & HD 192164 & 60 & 1.0 & $-11 \pm 1$ \\
& & & 55370 & 150 & 1.1 & HD 192164 & 120 & 1.0 & $-14 \pm 1$ \\
& & & 55371 & 450 & 1.1 & HD 192164 & 120 & 1.0 & $-10 \pm 1$ \\
& & & 55372 & 450 & 1.1 & HD 192164 & 60 & 1.0 & $-10 \pm 1$ \\
& & & 55695 & 450 & 1.3 & HD 192164 & 90 & 1.2  & $-9 \pm 1$ \\
& & & 55728 & 375 & 1.1 & HD 192164 & 100 & 1.0  & $-12 \pm 1$ \\
& & & 55729 & 375 & 1.1 & HD 192164 & 100 & 1.0  & $-12 \pm 1$ \\
\enddata
\tablecomments{Within each group (T Tauri stars, Herbig Ae stars,
  Herbig Be stars, FU Ori Sources, and heavily-veiled objects), 
  sources are listed roughly in order or right ascension,
  keeping targets that share common calibration in consecutive order.
  Uncertainties in measured EW include statistical uncertainties and
  uncertainties related to uncorrected bad pixels, but do
  not include errors in estimation of continuum level.}
\end{deluxetable*}

\end{document}